\documentclass{iau} 
\usepackage{graphicx}

\title[Complex dynamical evolution of supermassive black holes] %% give here short title %%
{The complex evolution of supermassive black holes in cosmological simulations}

\author[Peter H. Johansson et al.]   %% give here short author list %%
{Peter H. Johansson$^1$, Matias Mannerkoski$^1$, Antti Rantala$^2$, Shihong Liao$^1$, Alexander Rawlings$^1$, Dimitrios Irodotou$^1$, Francesco Rizzuto$^1$}

\affiliation{$^1$Department of Physics,
  Gustaf H\"allstr\"omin katu 2, FI-00014, University of Helsinki, Finland \\ email: {\tt Peter.Johansson@helsinki.fi} \\[\affilskip]
$^2$Max-Planck Institut f\"ur Astrophysik, Karl-Schwarzschild-Str 1, D-85748, Garching, Germany}

\pubyear{2022}
\volume{362}  %% insert here IAU Symposium No.
\jname{Predictive power of computational astrophysics as a discovery tool}
\editors{D. Bisikalo, C. Boily \& T. Hanawa, eds.}
\begin{document}

\maketitle

\begin{abstract}
  We present here self-consistent zoom-in simulations of massive galaxies forming in
  a full cosmological setting.  The simulations are run with an updated version of the KETJU code, which
  is able to resolve the  gravitational dynamics of their supermassive black holes, while simultaneously
  modelling the large-scale astrophysical processes in the surrounding galaxies, such as gas cooling, star formation and
  stellar and AGN feedback. The KETJU code is able to accurately model the complex behaviour of multiple SMBHs, including
  dynamical friction, stellar scattering and gravitational wave emission, and also to resolve Lidov--Kozai oscillations that
  naturally occur in hierarchical triplet SMBH systems. In general most of the SMBH binaries form at moderately high eccentricities,
  with typical values in the range of $e = 0.6-0.95$, meaning that the circular binary models that are commonly used in the literature are insufficient for capturing the typical binary evolution.

\end{abstract}

\firstsection % if your document starts with a section,
\section{Introduction}

In the $\Lambda$CDM model galaxies grow hierarchically through mergers and gas accretion (e.g. \cite[Naab \& Ostriker 2017]{Naab17}).
As all massive galaxies contain supermassive black holes (SMBHs) in their centres, the hierarchical growth of galaxies will invariably lead to
SMBH mergers, which typically proceed through a three-stage process (\cite[{Begelman} {et~al.} 1980]{begelman1980}). At large separations
the evolution of the SMBHs are driven by dynamical friction until a binary forms. In the next phase the SMBH binary
hardens through three-body scattering with individual stars (\cite[{{Hills} \& {Fullerton} 1980}]{hills1980}) and then finally at subparsec scales the binary
coalesces due to the emission of gravitational waves (\cite[{{Peters} 1964}]{1964PhRv..136.1224P}). 

Modelling this entire SMBH coalescence process in a full cosmological simulation has been very challenging due to the inability
of simultaneously modelling the small-scale SMBH dynamics and global galactic-scale astrophysical processes in simulations
that include gravitational softening (e.g. \cite[{Ryu} {et~al.} 2018]{2018MNRAS.473.3410R}). Instead, the parsec-scale dynamics has typically
been modelled by postprocessing the simulations using semi-analytic met-hods based on orbit-averaged equations (\cite[{Kelley} {et~al.} 2017]{2017MNRAS.464.3131K}) or by resimulating selected regions of galaxies by separate stand-alone N-body codes (\cite[{Khan} {et~al.} 2016]{2016ApJ...828...73K}). 

Here we present self-consistent 
cosmological zoom-in simulations
run with our updated KETJU code 
(\cite[{Rantala} {et~al.} 2017]{2017ApJ...840...53R}, \cite[{Rantala} {et~al.} 2018]{2018ApJ...864..113R}, \cite[{Mannerkoski} {et~al.} 2021]{2021ApJ...912L..20M}), which is able to resolve the dynamics of merging SMBHs down to tens of Schwarzschild radii, while
simultaneously modelling astrophysical processes in the surrounding galaxies, such as gas cooling, star formation and stellar and AGN
feedback. 

\section{Simulations}

In the KETJU code the dynamics of SMBHs and their surrounding stellar particles is integrated with the high-accuracy regularised integrator MSTAR (\cite[{Rantala} {et~al.} 2020]{2020MNRAS.492.4131R}), whereas the dynamics of the remaining particles is computed with the standard GADGET-3 leapfrog method (\cite[{{Springel} 2005}]{2005MNRAS.364.1105S}). The gravitational interactions of SMBHs with other SMBHs and stellar particles
are computed without softening while the interactions between stellar particles are softened in order to avoid energy errors when particles enter and exit the regularised KETJU region. The effects of general relativity, such as binary 
precession and gravitational wave (GW) emission are modelled by including post-Newtonian correction terms up to order 3.5 between each pair of SMBHs (\cite[{{Mora} \& {Will} 2004}]{2004PhRvD..69j4021M}). In addition, we also now include the 1PN corrections for general N-body systems, which could potentially affect the long-term evolution of triple and multiple SMBH systems (e.g. \cite[{{Will} 2014}]{2014PhRvD..89d4043W}).

The gas component is modelled using the SPHGAL smoothed particle hydrodynamics implementation 
(\cite[{Hu} {et~al.} 2014]{2014MNRAS.443.1173H}). We include metal-dependent gas cooling that tracks 11 individual elements and use a stochastic star formation model with a critical hydrogen number density threshold of $n_{\rm H}=0.1 \ \rm cm^{-3}$. The model also includes feedback from supernovae (both type II and Ia) and massive stars, as well as the production of metals through chemical evolution (\cite[{Aumer} {et~al.} 2013]{2013MNRAS.434.3142A}). Galaxies with dark matter halo masses of $M_{\rm DM}=10^{10} h^{-1} M_{\odot}$ are seeded with SMBHs with 
masses of $M_{\rm BH}=10^{5} h^{-1} M_{\odot}$, which first grow through standard Bondi--Hoyle--Lyttleton accretion and BH merging, with the 
maximum accretion rate capped at the Eddington limit, assuming a fixed radiative efficiency of $\epsilon_{r}=0.1$. A total of 0.5\% of the rest mass energy of the accreted gas is coupled to the surrounding gas as thermal feedback (\cite[{Johansson} {et~al.} 2009a]{2009ApJ...690..802J}). 

We run two cosmological zoom-in simulations starting at a redshift of $z=50$, with the initial conditions generated using the MUSIC software package (\cite[{{Hahn} \& {Abel} 2011}]{2011MNRAS.415.2101H}). The first 
simulation (simulation 1) targets a dark matter halo with a virial mass of $M_{\rm 200}\sim 7.5\times 10^{12} M_{\odot}$, whereas in the second simulation (simulation 2) we target a more massive system of  $M_{\rm 200}\sim 2.5\times 10^{13} M_{\odot}$, covering a larger initial comoving volume of $(10 h^{-1} \ \rm Mpc)^{3}$. The high-resolution zoom-in regions are initially populated with both gas and dark matter particles, with masses of 
$m_{\rm gas}=3\times 10^{5} \ M_{\odot}$ and  $m_{\rm DM}=1.6\times 10^{6} \ M_{\odot}$, respectively. The baryonic particles have gravitational 
softenings of $\epsilon_{\rm bar}=40 h^{-1} \ \rm pc$ for stars and gas and $\epsilon_{\rm DM}=93 h^{-1} \ \rm pc$ for the dark matter particles. The 
simulations are run initially with standard GADGET-3, until the SMBHs have grown to be sufficiently massive $(M_{\rm BH}\sim 7.5\times 10^{7} M_{\odot})$ to allow for detailed dynamical modelling using the KETJU code, as the algorithmically regularised integrator requires a BH to stellar particle mass ratio of $\sim 500-1000$ in order to provide accurate results (\cite[{Mannerkoski} {et~al.} 2019]{2019ApJ...887...35M}).

\section{Resolving SMBH triplet systems}

Simulation 1 was run with standard GADGET-3 until redshift $z\approx 0.62$. At this point the target halo hosted three massive galaxies (A, B and C), all containing their individual central SMBHs with masses in excess of $10^{8} M_{\odot}$ $(M_{\rm BH,A}=8.4\times 10^{8} M_{\odot}, M_{\rm BH,B}=1.1\times 10^{8} M_{\odot}$ and $M_{\rm BH,C}=2.1\times 10^{8} M_{\odot}$, see \cite[{Mannerkoski} {et~al.} 2021]{2021ApJ...912L..20M}). At
this stage we turned on the KETJU integration as the mass ratio between the SMBHs and the stellar particles was now sufficiently large. The radii of the
regularised KETJU regions were set to $120 h^{-1} \ \rm pc$, corresponding to three times the baryonic softening length. 

\begin{figure}[t]
% \vspace*{-2.0 cm}
\begin{center}
 \includegraphics[width=5.3in]{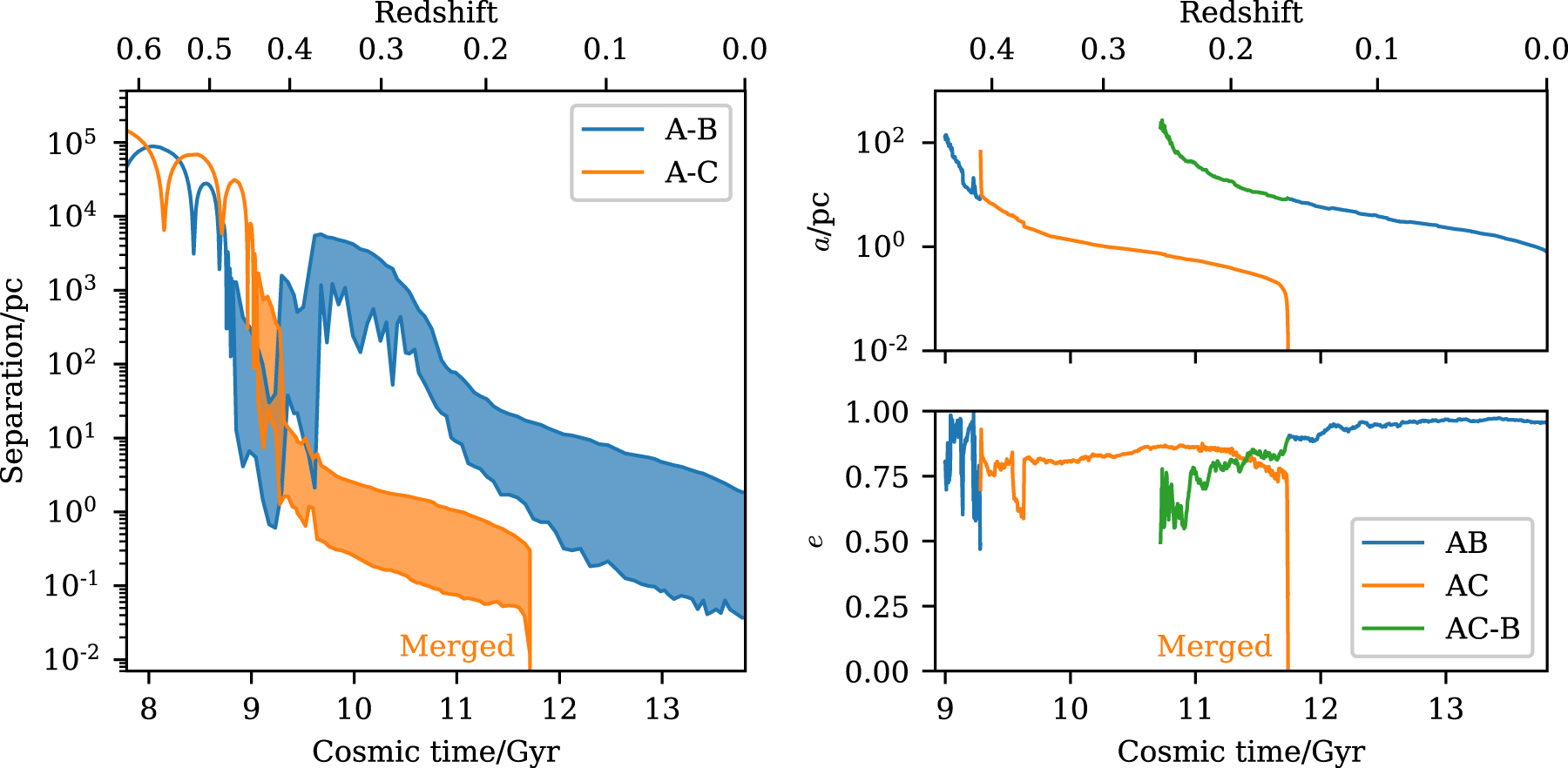} 
% \vspace*{-1.0 cm}
 \caption{Left: The separations of the A-B and A-C SMBHs over the duration of the KETJU simulation, with the shaded regions showing the 
range of rapid oscillations. Right: The evolution of the semimajor axis \textit{a} (top) and the eccentricity \textit{e} (bottom) for the SMBHs in the system (figure adapted from \cite[{Mannerkoski} {et~al.} 2021]{2021ApJ...912L..20M}).}
\label{Fig1}
\end{center}
\end{figure}

Galaxy B merges with galaxy A at a redshift of $z\approx 0.48$ and during the merger the two SMBHs sink towards the centre of the merger 
remnant forming a binary (AB-binary) with a semi-major axis of $a_{\rm AB}\approx 100 \ \rm pc$. This binary hardens through stellar scattering 
over the next $\sim 250 \ \rm Myr$ reaching a semi-major axis of  $a_{\rm AB}\approx 10 \ \rm pc$ (see Fig. \ref{Fig1}). However, before this 
binary enters into the gravitational wave dominated regime, galaxy C merges with the AB galaxy remnant bringing in SMBH-C in the process, which results in 
a three-body interaction between the three SMBHs. Initially, the three-body interaction causes rapid changes in the eccentricity 
of the AB-binary and finally SMBH-B is ejected from the centre, with SMBH-C instead replacing it in the new AC-binary. 

After a few hundred Myr, SMBH-B falls back towards the AC-binary resulting in an interaction with the AC-binary, which can be seen from the
small SMBH separations and the dip in the AC eccentricity in Fig. \ref{Fig1}. This interaction ejects SMBH-B to an even wider orbit, and it takes it 
around one Gyr to sink back into the centre.  In the meantime, the AC-binary hardens due to stellar scattering and finally merges driven 
by gravitational wave emission, roughly $\sim 3$ Gyr after the galaxies merged. The remaining AB-binary also hardens due to stellar scattering, 
but does not have time to merge before the simulation ends at $z=0$. 

The eccentricity of the AC-binary also exhibits small oscillations after SMBH-B enters into a sub $\sim 100 \ \rm pc$ hierarchical configuration. At
this stage the inner binary has a semi-major axis of  $a_{\rm AC}\approx 0.4 \ \rm pc$, while SMBH-B is on a much 
wider orbit with $a_{\rm AC-B}\approx 20 \ \rm pc$ and an eccentricity of $e_{\rm AC-B}\approx 0.79$ at an inclination of $i_{\rm AC-B}\approx 90.8^{\circ}$.
Here we are in fact witnessing Lidov--Kozai oscillations (\cite[{Lidov} 1962]{1962P&SS....9..719L}) suppressed by the relativistic precession of the inner orbit, due to the fact that the binary precession period $(\sim 6\times 10^{5} \ \rm yr)$ is much shorter than the Lidov--Kozai oscillation period $(\sim 4\times 10^{7} \ \rm yr)$ for this particular system (e.g. \cite[{Blaes} {et~al.} 2002]{2002ApJ...578..775B}).

\begin{figure}[t]
% \vspace*{-2.0 cm}
\begin{center}
 \includegraphics[width=5.3in]{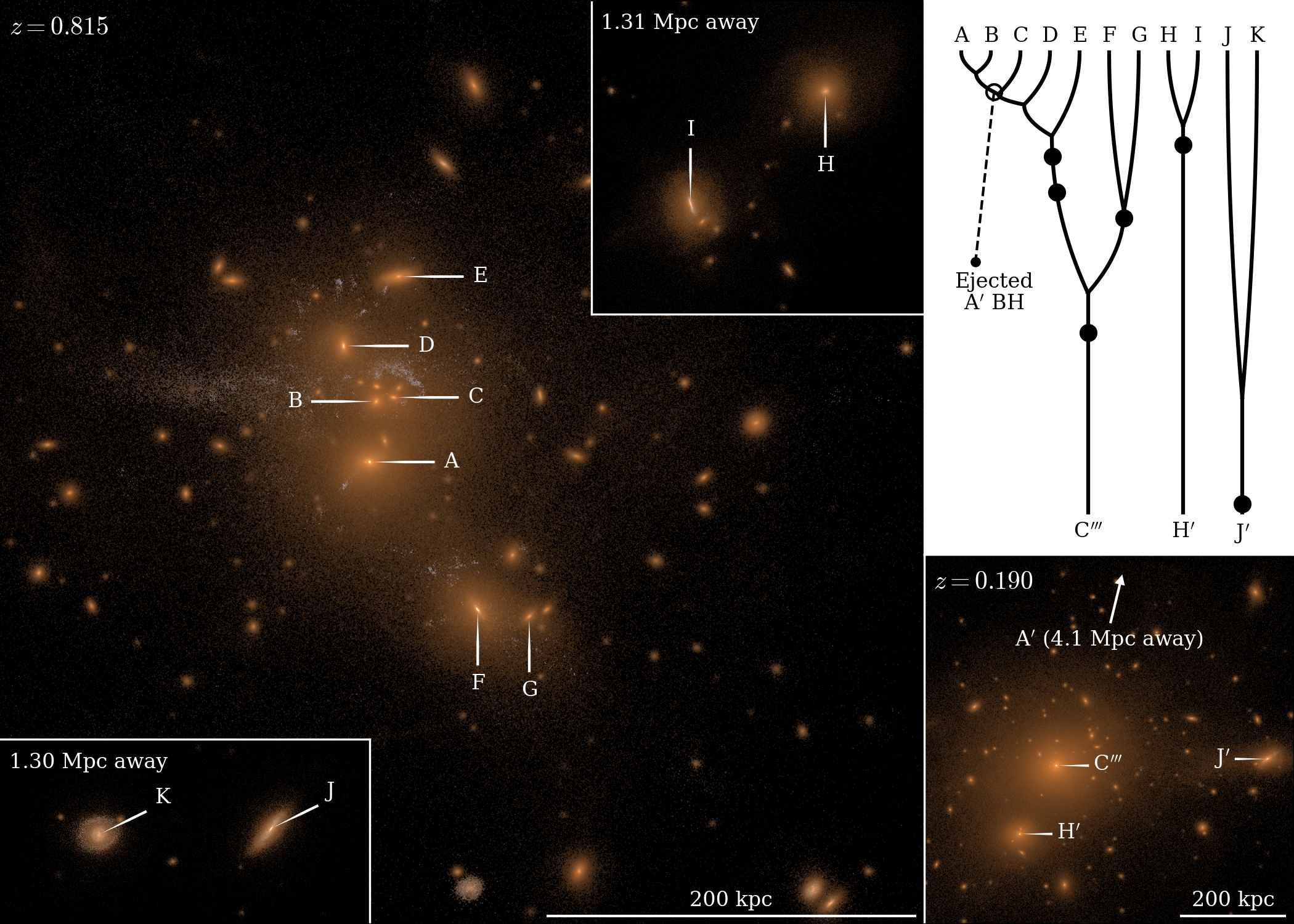} 
% \vspace*{-1.0 cm}
 \caption{Left: The initial state of the KETJU run, with the galaxies and SMBHs indicated. The main panel shows the central group of galaxies, with
two more distant galaxy pairs shown as insets in the corners. Top right: A schematic merger tree of the galaxies and their SMBHs, with time proceeding
from top to bottom. The lines depict galaxy mergers, while the circles indicate SMBH binary mergers. The final state of the KETJU simulation
at $z=0.190$ is shown in the bottom right corner, with the remaining SMBHs labelled (\cite[{Mannerkoski} {et~al.} 2022]{2021arXiv211203576M}).}
   \label{Fig2}
\end{center}
\end{figure}

\section{Simulating systems with multiple SMBHs}

In simulation 2 a larger comoving volume of $(10 h^{-1} \ \rm Mpc)^{3}$ was run initially with GADGET-3 until redshift $z\approx 0.815$, after which the integration
was continued with KETJU turned on (\cite[{Mannerkoski} {et~al.} 2022]{2021arXiv211203576M}). At the start of the KETJU simulation the volume contained 11 massive galaxies, with SMBHs that are resolved with their individual
regularised regions. The galaxies are shown in the left panel of Fig. \ref{Fig2} with seven galaxies (A-G), located in a central group that is collapsing within a halo with a total
virial mass of $M_{200}\approx 2\times 10^{13} M_{\odot}$. In addition, there are two more distant galaxy pairs, with H and I located in a halo with a virial mass
of $M_{200}\approx 2.5\times 10^{12} M_{\odot}$, and K and J found in a halo with a virial mass of $M_{200}\approx 1.3\times 10^{12} M_{\odot}$. Due to the high number of massive
black holes in this simulation, we lowered the gravitational softening to $\epsilon_{\star}=20 h^{-1} \ \rm pc$ for the KETJU simulation, which allowed us to resolve regularised
regions around each SMBH with a radius of $60 h^{-1} \ \rm pc$.

The galaxies and their constituent SMBHs undergo multiple mergers during the KETJU simulation, which is depicted schematically in the top right panel of Fig. \ref{Fig2}. In this
simulation we also include a description for SMBH spins and model their gravitational wave driven merger kicks using an analytic model based on numerical relativity fitting functions
from \cite[{{Zlochower} \& {Lousto} (2015)}]{2015PhRvD..92b4022Z}. Typically the SMBH merger remnants experience rather modest kicks of $v_{\rm kick}\lesssim 500 \ \rm km/s$, the exception being the AB-SMBH remnant, which receives a very large kick of  $v_{\rm kick}= 2257 \ \rm km/s$, which is sufficient to eject the SMBH from its host galaxy. Thus, galaxy
A is temporarily lacking a SMBH, however this situation is rapidly remedied with the subsequent mergers of galaxies C, D and E, which bring in their central SMBHs replacing the ejected
SMBH. The fact that the original SMBH was ejected from this galaxy has important consequences for the evolution of the galaxy on the $M_{\rm BH}- \sigma$ plane, as the galaxy will have
an undermassive SMBH with respect to the observed relation (\cite[{Johansson} {et~al.} 2009b]{2009ApJ...707L.184J}, \cite[{{Kormendy} \& {Ho} 2013}]{kormendy2013}, see also \cite[{Mannerkoski} {et~al.} 2022]{2021arXiv211203576M} for details).  

\begin{figure}[t]
% \vspace*{-2.0 cm}
\begin{center}
 \includegraphics[width=5.3in]{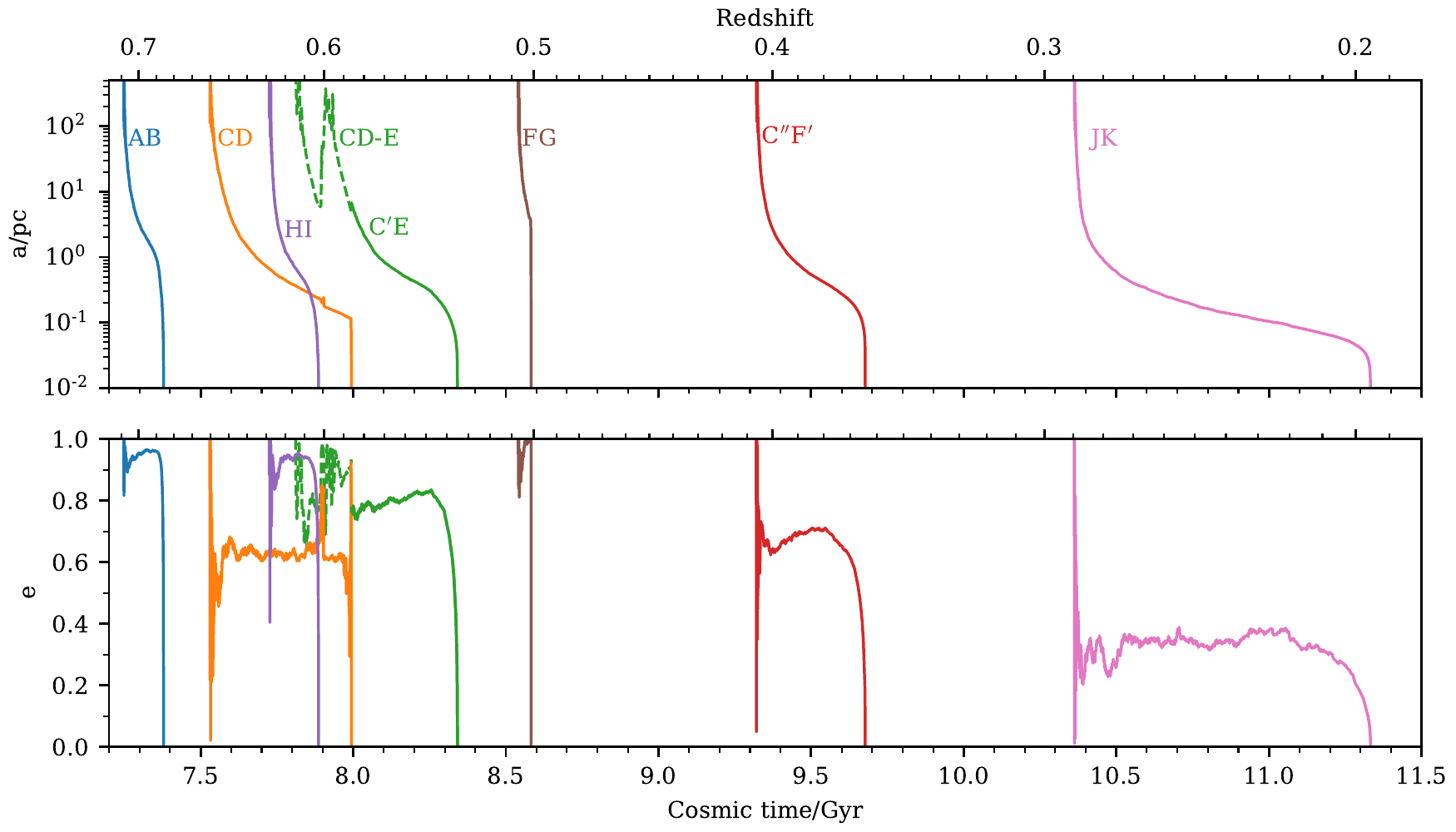} 
% \vspace*{-1.0 cm}
 \caption{The evolution of the semimajor axis $a$ (top) and the eccentricity $e$ (bottom) of the SMBH binaries in simulation 2. The dashed line shows the parameters of the outer orbit in the hierarchical CD-E triplet system. The binaries form with eccentricities in a broad range of $e\sim 0.3-0.9$ (figure adapted from \cite[{Mannerkoski} {et~al.} 2022]{2021arXiv211203576M}).}
   \label{Fig3}
\end{center}
\end{figure}

In Fig. \ref{Fig3} we show the evolution of the semi-major axis and eccentricity for all the resolved massive SMBH mergers in the simulation as a function of redshift. In general most of the SMBH binaries form at moderately high eccentricities, with typical values in the range of $e=0.6-0.95$ and limited eccentricity evolution during the hardening process. The relatively high eccentricities result in short binary lifetimes with the SMBH coalescense typically occurring within $\sim 200-500 \ \rm Myr$. However, there are some notable exceptions, for example the FG-binary has an extremely high eccentricity of $e=0.998$, which results in a very rapid gravitational wave driven merger within just a few tens of Myr. For this binary
most of the eccentricity growth occurs when the binary semimajor axis is still above $\sim 10 \ \rm pc$, and the mass ratio of the binary is large $(\sim 7:1)$, implying that resonant
dynamical friction (\cite[{{Rauch} \& {Tremaine} 1996}]{1996NewA....1..149R}) might also be operational, in addition to simple stellar scattering (\cite[{{Quinlan} 1996}]{1996NewA....1...35Q}).

The JK-binary on the other hand has a low eccentricity of only $e=0.35$, and is formed after a nearly circular orbit galaxy merger. The low eccentricity results in a slow merger process and it takes nearly a Gyr for the black holes to merge after forming a hard binary. Finally, similarly to simulation 1, a SMBH triplet (CD-E) is also occurring in this simulation (Fig. \ref{Fig3}). After a strong gravitational interaction with the CD-binary, SMBH-E settles into a hierarchical triplet configuration around the inner binary. However, contrary to SMBH triplet in simulation 1, the outer period is in this case shorter than the relativistic period of the inner binary. This results in Lidov--Kozai oscillations (\cite[{Lidov} 1962]{1962P&SS....9..719L}) that eventually excite the CD-binary eccentricity from $e\approx 0.55$ to a very high value of $e\approx 0.9$, and the increased eccentricity is sufficient
to drive the CD-binary to a near instant merger through the increased emission of gravitational waves.

\section{Conclusions}

We have demonstrated here that the KETJU code can be used to resolve the detailed small-scale dynamics of tens of SMBHs evolving in a complex cosmological environment over extended periods of time. All SMBH binary systems found in our simulations were driven to merger by stellar interactions without any signs of stalling. Our simulated binaries typically formed
on highly eccentric orbits, indicating that the circular binary models that are commonly used in the literature are insufficient for capturing the typical binary evolution. In addition, we found that systems with multiple interacting SMBHs naturally occur in a $\Lambda$CDM setting and it is important to capture their dynamics accurately, which can only be done with
direct integrations of the type presented here. Finally, we stress the importance of simultaneously modelling the accurate small-scale SMBH dynamics and gas dynamics, which will be
in particular important when making gravitational wave predictions for LISA (\cite[{Amaro-Seoane} {et~al.} 2022]{2022arXiv220306016A}), as it will be mostly sensitive to somewhat lower-mass SMBHs in the mass range of $M_{\rm BH}\sim 10^{5}-10^{7} M_{\odot}$, which are expected to reside in late-type gas-rich galaxies.  

\section*{Acknowledgments}

The authors acknowledge the support by the ERC via Consolidator Grant KETJU (no. 818930) and the support of the Academy of Finland grant 339127.

\end{document}